\documentclass[a4paper,fleqn,usenatbib]{mnras}

\usepackage[T1]{fontenc}
\usepackage{ae,aecompl}

\usepackage{graphicx}	
\usepackage{amsmath}	
\usepackage{amssymb}	

\title[Period modulation in BE Dor]{Reanalysis of c-type RR Lyrae Variable BE Dor, Period Modulations and Possible Mechanism}
\author[L. -J. Li et al.]{
L. -J. Li,$^{1}$\thanks{E-mail: lipk@ynao.ac.cn}
S. -B. Qian,$^{1,2}$
and L. -Y. Zhu$^{1,2}$
\\
$^1$Yunnan Observatories, Chinese Academy of Sciences, P.O.Box110, Kunming 650011, Yunnan, P.R.China \\
$^2$University of Chinese Academy of Sciences, No.19(A) Yuquan Road, Shijingshan District, Beijing 100049, P.R.China\\}
\date{Accepted XXX. Received YYY; in original form ZZZ}

\pubyear{2016}

\begin{document}
\label{firstpage}
\pagerange{\pageref{firstpage}--\pageref{lastpage}}
\maketitle

\begin{abstract}

We reanalyzed the c-type RR Lyrae star BE Dor (MACHO 5.4644.8, OGLE-LMC-RRLYR-06002) that had been discovered to show cyclic period changes. The photometric data of several sky surveys (DASCH, MACHO, OGLE, ASAS-SN, and TESS) were used for analyses. The $O-C$ diagram and pulsation period obtained from Fourier analysis show significant period modulations in BE Dor. However, different from the previous viewpoint, the changes are quasi-periodic and abrupt. Therefore, the light-travel time effect caused by the companion motion cannot explain the changes. Noting a same subtype star KIC 9453114 with similar phenomena has a high macroturbulent velocity, and the degree of $O-C$ changes seem to be positively correlated with these velocities, we consider that the mechanism leading to period modulation should be caused by the interaction between turbulent convection and magnetic field activity in the ionization zone, i.e., the viewpoint of Stothers. It may not explain the general Blazhko effect but should explain such period modulations in BE Dor and those other c-type RR Lyrae stars. We hope our discoveries and viewpoints can provide some information and inspiration for relevant research.

\end{abstract}

\begin{keywords}
techniques: photometric --
stars: fundamental parameters --
stars: variables : RR Lyrae
stars: individual : BE Dor
\end{keywords}



\section{Introduction} \label{sec:intro}

RR Lyrae stars (RRLs) are a kind of short pulsation period variable stars widely distributed in various celestial systems in the Galaxy. According to the characteristics of the light curves and pulsation modes, RRLs can be classified as ab-type RRLs (fundamental mode, abbreviated as RRab), c-type RRLs (first overtone mode, RRc), and d-type RRLs (double modes, RRd). Because of their characteristics such as period-luminosity relation, RRLs are usually used as probes to study the dynamics and chemical evolution of old age low mass stars in the Milky Way \citep{2004rrls.book.....S}.

In the field of RRLs, the relevant photometric observations and researches are mostly carried out around their pulsations. The pulsations are affected by different factors. Therefore, the researchers can use the pulsations as probes to explore the mechanism behind them. The pulsation variations of long-term scale (decades or even hundreds of years) are generally considered to be caused by stellar evolutions and light-travel time effect (\citealt{2021AJ....161..193L} and references therein). While the short-term variations are usually known as Blazhko effect, defined as the periodic or quasi-periodic variations of the pulsation parameters (i.e., pulsation period, amplitude, and phase) with time \citep{1907AN....175..325B}. With the running of multiple sky survey projects, a tremendous amount of photometric data have accumulated, and hundreds of Blazhko star candidates have been obtained (\citealt{2013A&A...549A.101S,2018MNRAS.480.1229N}, and references therein). Recently, with the launch of space projects (e.g. $CoRoT$, $Kepler$), researchers have obtained unprecedented high-precision uninterrupted data, and great progress has been made in the discoveries and researches of Blazhko stars \citep{2009AIPC.1170..235C,2010A&A...510A..39C,2010A&A...520A.108P,2011A&A...527A.146C,2013arXiv1310.0542K}.

Compared with the research on the RRab stars, there are relatively fewer studies on RRc stars. Recently, \citet{2013JAVSO..41...75P} studied the period changes of 40 RRc stars by using the $O-C$ method, and pointed out that the linear period changes of most targets are consistent with the prediction of the stellar evolution model. Some RRc stars are also found to have special characteristics, e.g., existing two close pulsation frequencies \citep{2005IBVS.5632....1A,2012PZP....12...18K}, or additional non-radial pulsation \citep{2015PZ.....35....5K,2015MNRAS.447.2348M}. In addition, it is also noted that the pulsations of some RRc stars have significant period modulations/jumps, but in contrast, the pulsation amplitude has no significant corresponding variation \citep{1981A&A...103..339O,2004MNRAS.354..821D,2007IBVS.5765....1W,2008PZ.....28....1W,2017IBVS.6212....1B,2019arXiv190200905L,2021AcA....71...55B}.

BE Dor ($\alpha_{2000} = 05^{\rm h}:09^{\rm m}:18^{\rm s}.94$, $\delta_{2000} = -69^{\circ}:50^{'}:14^{''}.43$, $\langle V \rangle$ = 15.231 mag, $\langle I \rangle$ = 14.861 mag, $\Delta I$ = 0.302 mag, MACHO 5.4644.8, OGLE LMC-RRLYR-06002) is a foreground RRc star toward the Large Magellanic Cloud (LMC). It was listed for the first time by \citet{1997ApJ...490L..59A} as an RRc star. In that paper, BE Dor, together with other LMC foreground RRLs, were used as tracers to find the possible intervening galaxy in front of the LMC. At that time, the period change of BE Dor has been noticed. A further detailed study was made by \citet{2004MNRAS.354..821D}: they carried out $O-C$ analysis, period analysis, and Fourier spectrum analysis on BE Dor by using MACHO and OGLE-III photometric data; found that the pulsation period has a periodic change with a period of about 7.95 years, but other pulsation parameters and Fourier coefficients have no obvious change, indicating that there is only simple period modulation in BE Dor. \citet{2004MNRAS.354..821D} provided two explanations for the relevant physical mechanism: one is the light-travel time effect \citep{1952ApJ...116..211I}, and the second is magnetic activity cycles \citep{1980PASP...92..475S}.

In the present paper, we collect and organize photometric data from several sky survey projects, reanalyze the pulsation of BE Dor by using the $O-C$ method and Fourier analysis method, study its period changes on a longer time baseline, and try to confirm the mechanism behind the phenomenon. Section~\ref{sec:surveys} introduces the photometric data from different sky surveys, including the properties of data, the description of data processing, and analysis. Section~\ref{Sec:Analyses} is the corresponding analyses, and in Section~\ref{Sec:Discussion} and~\ref{Sec:Summary}, the discussions and summaries are presented, respectively.

\section{Sky Surveys} \label{sec:surveys}

BE Dor locates in the southern sky region, and we mainly use the data provided by those photometric sky survey projects. After searching, we find that the photographic plate digitization project DASCH\footnote{\url{https://projects.iq.harvard.edu/dasch}} (Digital Access to a sky century @ Harvard, \citealt{2009ASPC..410..101G}), the space sky survey project TESS\footnote{\url{https://archive.stsci.edu/missions-and-data/tess}} (Transiting Exoplanet Survey Satellite, \citealt{2015JATIS...1a4003R}), and ASAS-SN\footnote{\url{http://www.astronomy.ohio-state.edu/asassn}} (All-Sky Automated Survey for Supernovae, \citealt{2014ApJ...788...48S}) provide the relevant data of BE Dor. Of course, the MACHO\footnote{\url{https://wwwmacho.anu.edu.au/}} (MAssive Compact Halo Objects, \citealt{1997ApJ...486..697A}) and OGLE\footnote{\url{http://ogle.astrouw.edu.pl/}} (The Optical Gravitational Lensing Experiment, \citealt{1992AcA....42..253U}) data are also indispensable. We will introduce them respectively below.

\subsection{DASCH data} \label{sec:surveys:DASCH}

DASCH is a project to convert the photographic plate data accumulated by different telescopes around the world in the early years into digital data \citep{2009ASPC..410..101G}. Their characteristics are low time resolution and low data precision (about 0.1 mag), but the advantage is long time baseline. The span of the light curves of some variable stars can reach decades or even 100 years. Therefore, by selecting appropriate variables star targets (e.g. period $>$ 0.2 days, amplitude $>$ 0.2 mag) and adopting suitable processing methods, we can obtain credible and valuable information. Actually, in recent years, researchers have published a series of works by using DASCH data \citep{2016A&A...589A..94L,2016MNRAS.459.4360L,2018MNRAS.474..824S,2018PASJ...70...71L,2018ApJ...863..151L,2021AJ....161..193L}.

\begin{figure}
\centering
\includegraphics[width=.45\textwidth]{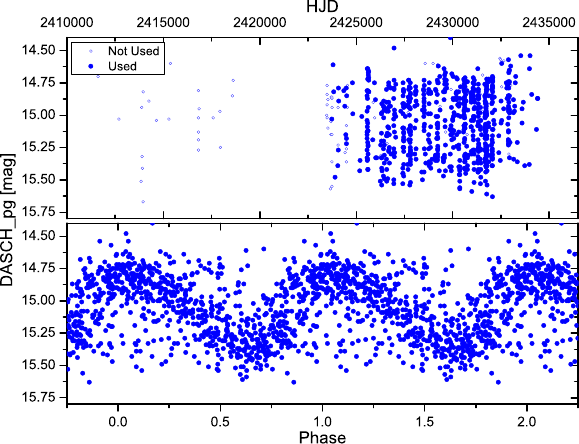}
\caption{Light curves of BE Dor provided by DASCH. The photograph magnitude is close to B-band magnitude. In the top panel, empty blue circles denote those data not used in the analyses. The phased light curves are plotted in the bottom panel, and the linear ephemeris HJD$_{\rm MAX}$ = 2430118.2192 + 0$^{\rm d}$.3279975$\cdot E$ is used to calculate the phases.} \label{Fig.1}
\end{figure}

DASCH provides about 800 data points of BE Dor lasting from 1890 to 1953, of which there are only 23 data points from 1890 to 1909. Due to the low time resolution, these 23 points were not used in the analysis. The exposure times of most observations are 45 or 60 minutes, and only a few are greater than 60 minutes. Therefore, we finally use 720 data points with exposure times less than 60 minutes from 1923 to 1953 for analysis. Figure~\ref{Fig.1} presents the DASCH light curves of BE Dor in the top panel, and the bottom panel shows the phased light curves. Due to the low time resolution, the DASCH data are mainly used for $O-C$ analysis.

We adopt an unconventional method to determine the times of light maximum, which has successfully applied in previous works \citep{2018ApJ...863..151L,2018PASJ...70...71L,2021AJ....161..193L}. We first divide the light curves into multiple segments; fit each segment with the following Fourier polynomials,
\begin{equation}
m(t)=A_{0}+\sum^{n}_{k=1}A_{k}\sin[\frac{2\pi k t}{P_{\rm pul}}+\phi^{\rm s}_{k}], \label{Equ:1}
\end{equation}
where $m(t)$ is the magnitude observed at time $t$, $A_{0}$ is the mean magnitude, $A_{k}$ is the amplitude of the $k$ component, and $\phi^{\rm s}_{k}$ is the phase of the $k$ component in $\sin$ term; then obtained the times of light maximum according to the fitting results. Considering that the light curves of RRc stars are approximately sinusoidal, and taking higher terms cannot get reliable results, we set the term $n$ as 1 in the fitting. Another problem that should be considered in the analysis is the maximum time delay \citep{2016A&A...589A..94L,2021AJ....161..193L}, that is, the obtained times of light maximum will be later than the real times due to long exposure times and low time resolutions. Similar to previous work \citep{2021AJ....161..193L}, we took KIC 4064484 as the template (pulsation period is about 0.337 days), and carried out experiments with $Kepler$ SLC data to correct the delay effect. We found that when $n$=1, the times of light maximum obtained is systematically 10 minutes later than the real times, and are insensitive to the exposure times. Table~\ref{Table1} lists the corresponding corrected times of light maximum. In addition, it can be seen from Equation~(\ref{Equ:1}) that we also obtain the mean magnitude $A_{0}$, pulsation period $P_{\rm pul}$ and other parameters of each data segment.

\subsection{MACHO and OGLE} \label{sec:surveys:MACHOOGLE}

The main scientific objectives of MACHO and OGLE were to find dark matter in the universe by microlensing technique \citep{1992AcA....42..253U,1997ApJ...486..697A}. They respectively monitored those regions, such as the Galactic Bulge, the large and small Magellanic Clouds. In addition to many achievements in the discovery of gravitational lens events, these two projects have also brought results in the field of variable stars (e.g., RRLs in LMC, \citealt{1996AJ....111.1146A,2003AcA....53...93S,2009AcA....59....1S,2016AcA....66..131S,2019AcA....69...87S}). At present MACHO project has been completed, but OGLE project is still ongoing and has reached phase IV. We collected these old and new data for the $O-C$ and Fourier analyses, and compared the results with the work of \citet{2004MNRAS.354..821D}.

\subsection{TESS} \label{sec:surveys:TESS}

TESS is a new space project after the $Kepler$ project, which also uses the transit method to discover exoplanets in the nearby stars \citep{2015JATIS...1a4003R}. The advantage of TESS is its observation region can basically cover the whole celestial surface. However, the TESS telescope has a small aperture, and its resolution is only 21 arcseconds/pixel. Therefore, TESS cannot directly provide the light curves for faint stars in dense star fields such as BE Dor. Fortunately, TESS also provides the download of cutouts of images \citep{2019ascl.soft05007B}, which enables us to try targeted processing of data. Until July 24, 2021, TESS provides the image files of 25 sectors for BE Dor (22 files contain valid data), and we download the corresponding $5\times5$ pixel cutouts. Through preliminary observation, we find that the fluxes of the central pixels change periodically, which indicates that the images do carry the light information of BE Dor. Further, we plot the light curves of the 25 pixels over time independently and find that the flux variations caused by BE Dor concentrate in the center $3\times3$ pixels.

\begin{figure}
\centering
\includegraphics[width=.45\textwidth]{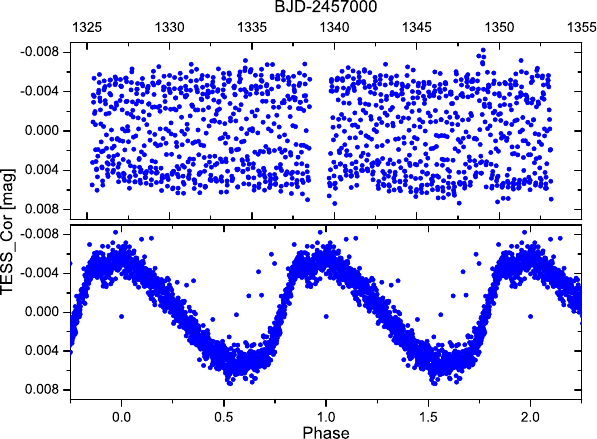}
\caption{Part of light curves of BE Dor obtained from TESS (Sector 1, Camera 4, CCD1). The phases in the bottom panel are calculated by the linear ephemeris HJD$_{\rm MAX}$ = 2458335.9211 + 0$^{\rm d}$.3280228$\cdot E$. It should be noted that the vertical axis is just on a magnitude scale. The intrinsic amplitude is heavily decreased due to the contamination.} \label{Fig.2}
\end{figure}

Therefore, we assume that there is only one variable star (i.e. BE Dor) in the whole $5\times5$ pixel image, the light variations are only concentrated in the central $3\times3$ pixel image, and the 16 pixels in the periphery do not contain the information of BE Dor. To eliminate the flux changes caused by instrument and other factors, the following equation is used to obtain the light curves:
\begin{equation}
\Delta mag(t) = -2.5 \lg[\frac{\sum^{9}_{\rm inner=1} flux\_inner(t)}{\sum^{16}_{\rm outer=1}flux\_outer(t)}\frac{16}{9}]
\end{equation}
In the above formula, other non-pulsation variation factors are well eliminated by the division. After obtaining the light curves of each sector (most span 30 days), we fit a polynomial to the data and then remove the long-term trends by subtracting the polynomial curve. Figure~\ref{Fig.2} shows the light curves after treatments. Due to the low resolution, BE Dor cannot be distinguished from the background stars, and the amplitude of light curves in Figure~\ref{Fig.2} is far less than the real amplitude of BE Dor. According to the full amplitude of the obtained light curves (0.013 - 0.017 mag) and the known full amplitude ($\Delta I$ = 0.302 mag), we estimate that the fluxes provided by BE Dor are 4.4 - 5.7\% of total $3\times3$ pixel fluxes.

\subsection{ASAS-SN} \label{sec:surveys:ASASSN}

\begin{figure}
\centering
\includegraphics[width=.45\textwidth]{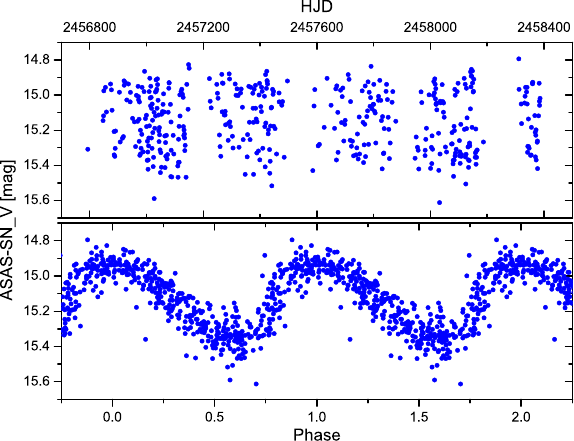}
\caption{V-band light curves of BE Dor provided by ASAS-SN. The phases in the bottom panel are calculated by the linear ephemeris HJD$_{\rm MAX}$ = 2457732.35095 + 0$^{\rm d}$.32802602$\cdot E$.} \label{Fig.3}
\end{figure}

ASAS-SN project uses multiple telescopes located in different countries to monitor the vast sky with the purpose of discovering bright supernovas and other transients \citep{2014ApJ...788...48S}. Meanwhile, it also obtains the light curves of a large number of variable stars, and sorts out the catalogs \citep{2018MNRAS.477.3145J,2020yCat.2366....0J}. Its variable stars database provides five years of V-band light curves for BE Dor, which are similar to MACHO and OGLE data in time resolution and precision. Figure~\ref{Fig.3} shows the corresponding light curves of BE Dor.

\section{Analyses} \label{Sec:Analyses}

The apparent magnitude of BE Dor is only $\langle V \rangle$ $\sim$ 15 mag, most of the data provided by sky survey projects are obtained by photometric observation. Therefore, our analyses focus on these photometric data. The corresponding methods are the $O-C$ method and the Fourier analysis method. The analyses are presented in the following subsections.

\subsection{$O-C$ analysis}
\begin{figure}
\includegraphics[width=.45\textwidth]{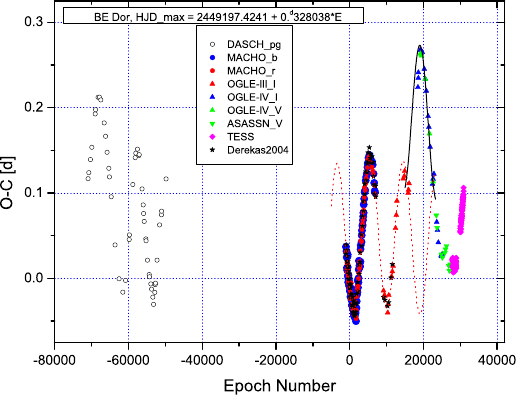}
\caption{$O-C$ diagram of BE Dor. The $O-C$ points show quasi-periodic variation and sudden jump change.} \label{Fig.4}
\end{figure}

For pulsating variable stars, $O-C$ analysis usually selects the light maximum as the reference phase to calculate the $O-C$ values \citep{2005ASPC..335....3S}. We have briefly introduced the method of determining the times of light maximum of BE Dor in Section~\ref{sec:surveys:DASCH}. Based on the different characteristics of photometric data, we set different Fourier orders when calculating the times of light maximum: for DASCH data, $n$ = 1; for MACHO, OGLE and ASAS-SN data, $n$ = 3; for TESS data, $n$ = 5. Table~\ref{Table1} lists the corresponding times of light maximum. Using the following corrected linear ephemeris,
\begin{equation}
\rm HJD_{\rm max} = 2449197.4241 + 0.^{d}328038\cdot E,
\end{equation}
we calculate the $O-C$ values and plot the $O-C$ diagram (see Figure~\ref{Fig.4}). In Figure~\ref{Fig.4}, the $O-C$ values provided by \citet{2004MNRAS.354..821D} are also plotted (black pentagram). It can be seen that the change trends of $O-C$ points obtained by us and \citet{2004MNRAS.354..821D} according to MACHO and OGLE-III data are consistent, and there is no obvious systematic deviation between them.

It is easy to see that the $O-C$ diagram of BE Dor shows more complex changes than the previous results. Firstly, when epoch number $<$ 15000 cycle, the $O-C$ curve shows a relatively periodic change, but the light-travel time effect model cannot well describe the changing trend. It means that the change is not strictly periodic. We note that there is a long gap between DASCH and MACHO data in the $O-C$ diagram (range from -50000 to -5000 epoch cycles). Since the $O-C$ curve of BE Dor fluctuates violently, it cannot rule out the epoch cycles of DASCH data are incorrect. Therefore, we abandoned the DASCH data, and only fitted the $O-C$ data of MACHO and OGLE-III. It is found that the $O-C$ diagram can be well described by the sinusoidal component with a period of about 8.1 years (red dotted line). According to the above result, it can be seen that the $O-C$ values suddenly changed the trend near the 15000 cycles, and the $O-C$ values should be reduced becomes increased. In other words, it seems that there is a $\pi$ phase jump around the 15000 cycles (black solid line refers to the $\pi$ phase jump). After that, the $O-C$ changes obtained by OGLE-IV, ASAS-SN, and TESS are also different from the previous change trend. Considering that the light-travel time effect model will generally cause the $O-C$ diagram to show strict periodic changes, it can not explain the $O-C$ diagram of BE Dor anymore.

\begin{figure}
\includegraphics[width=.45\textwidth]{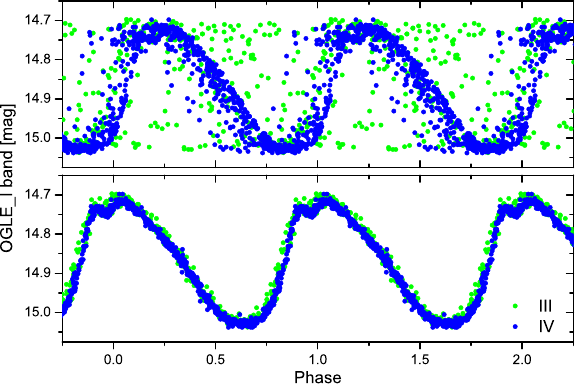}
\caption{The phased diagrams for BE Dor. The I-band data provided by OGLE-III (green) and IV (blue) are used.} \label{Fig.5}
\end{figure}

\citet{2004MNRAS.354..821D} plotted the phase diagrams after eliminating the phase shifts (see Figure 5 in their paper) and pointed out that there is no evidence for any additional variation in the phased light curves. We adopt a similar method and plot the phased diagrams in Figure~\ref{Fig.5} by using OGLE I-band data. In the upper panel, the phases of light curves are calculated from a linear ephemeris ($2455301.7898 + 0.^{\rm d}3279969 \cdot E$). The diagram shows the strong period/phase modulations, which are also shown in Figure 1 of \citet{2004MNRAS.354..821D}. The bottom panel shows the phase diagram after eliminating the modulation component according to our $O-C$ diagram, from which it can be seen that other pulsation parameters (e.g. full amplitude) have no obvious change. This confirms the viewpoint of \citet{2004MNRAS.354..821D}.

\subsection{Fourier analysis}

The changes in the $O-C$ diagram mean that the pulsation of BE Dor is period modulated. To explore whether there are corresponding changes in other pulsation parameters, we adopted Fourier analysis method to analyze the MACHO, OGLE, and TESS data (due to the large errors, the results of ASAS-SN data are not listed). The software used here is Period04 \citep{2004IAUS..224..786L,2005CoAst.146...53L}. Due to the strong period modulation, we cannot simply Fourier decompose the light curves as a whole but divide the light curves into many segments, and then Fourier analyzes each segment. The Fourier formula is the same as Equation~(\ref{Equ:1}). Table~\ref{Table2} lists the corresponding Fourier coefficients obtained from each segments, including pulsation period $P_{\rm pul}$, mean magnitude $A_{0}$, $R_{21}$, $R_{31}$, $\phi_{21}$, $\phi_{31}$ and corresponding errors.

\begin{figure}
\centering
\includegraphics[width=.45\textwidth]{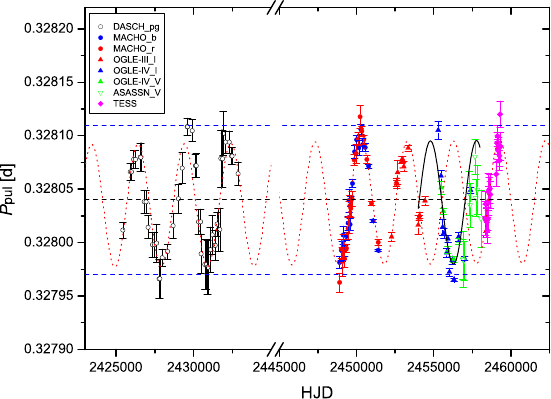}
\caption{$P_{\rm pul}$ vs Time diagram. The range of the period change is about $1.4\times10^{-4}$ days, i.e., 12.1 s.} \label{Fig.6}
\end{figure}

Figure~\ref{Fig.6} describes the trend chart of pulsation period with time, in which the black and green hollow points represent the pulsation period obtained from DASCH and ASAS-SN data, respectively. These values are obtained during the process of determining the times of light maximum. It can be seen that at first, the pulsation period changes quasi-periodically with time. The corresponding period change is about 12 s, which is consistent with the previous results \citep{2004MNRAS.354..821D}. The red dotted line indicates the corresponding sinusoidal fitting (only those $P_{\rm pul}$ determined from DASCH, MACHO, OGLE-III data are used), and the sinusoidal period is about 8.2 yr. However, between OGLE-III and IV, the pulsation period changes suddenly (black solid line refers to the $\pi$ phase jump). These results are consistent with those in the $O-C$ diagram. In addition, it can be seen from Figure~\ref{Fig.6} that the variation of $P_{\rm pul}$ determined from TESS (magenta diamonds) is consistent with the sinusoidal curve (red dotted line), although they are not used in the fitting just now.

\begin{figure*}
\centering
\includegraphics[width=.45\textwidth]{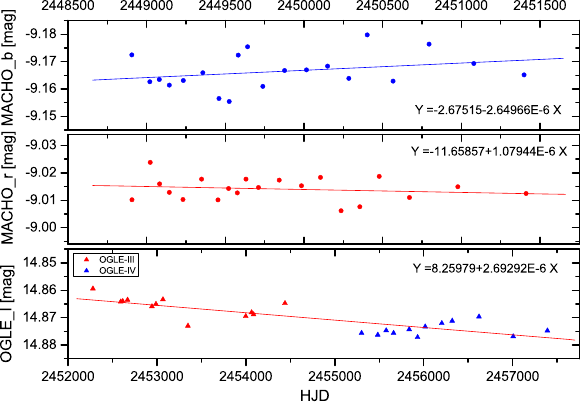}
\caption{Mean magnitude $A_{0}$ vs Time diagrams. The mean magnitudes of different bands seem to increase or decrease with time.} \label{Fig.7}
\end{figure*}

In addition, Figure~\ref{Fig.7} plots the variation of mean magnitudes with time. It can be seen that the mean magnitudes obtained from OGLE I-band data seem to increase with time (bottom panel), indicating that the luminosity in I-band is decreased. And accordingly, the values from the MACHO b-band are reduced with time (upper panel), but those from MACHO r-band data do not change significantly (middle panel). The effective wavelength range of I-band used in OGLE is 700 - 1100 nm, those of r and b filters used in MACHO are 590 - 780 nm and 440 - 590 nm, respectively. Assuming that the mean magnitudes do change as shown in Figure~\ref{Fig.7}, it means that the luminosity in the shortwave band becomes brighter and in the longwave band becomes darker. That is, the color index (b - I) decreases, and the effective temperature increases with time. Of course, this change is still very small, which is equivalent to the photometric error of the two projects. The long-term change of the magnitudes cannot rule out to be caused by instruments or other factors. In fact, for the photometric data accumulated for many years, how to retain the information of the long-term changes as much as possible in data processing and analysis is still a difficult problem. Therefore, we doubt the authenticity of the long-term changes of BE Dor mean magnitudes, and just make a presentation in Figure~\ref{Fig.7}.

\begin{figure*}
\centering
\includegraphics[width=.7\textwidth]{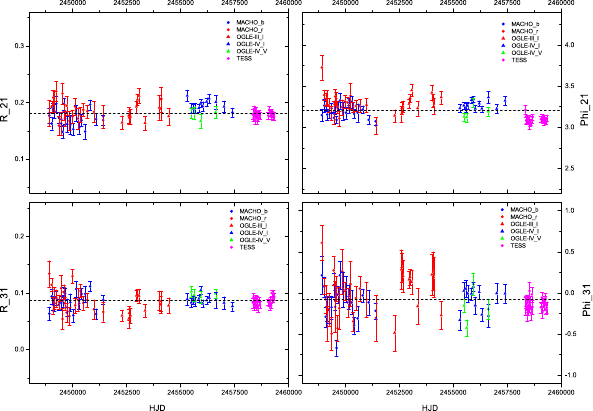}
\caption{Variations of the Fourier coefficients for BE Dor.} \label{Fig.8}
\end{figure*}

Figure~\ref{Fig.8} presents the diagrams of Fourier coefficients with time. It can be seen that these parameters fluctuate with time, but the changes are not so significant relative to the pulsation period (see Figure~\ref{Fig.6}). Of course, the Fourier coefficients obtained from different projects are in different passbands. There are systematic weak deviations between them. However, our coefficient errors are on the order of 10$^{-1}$ - 10$^{-2}$, which are greater than or equivalent to the deviations caused by passbands. Therefore, for the sake of brevity, we plot the coefficients in the same diagrams and mark them with different colors and graphics.

\begin{figure*}
\centering
\includegraphics[width=.45\textwidth]{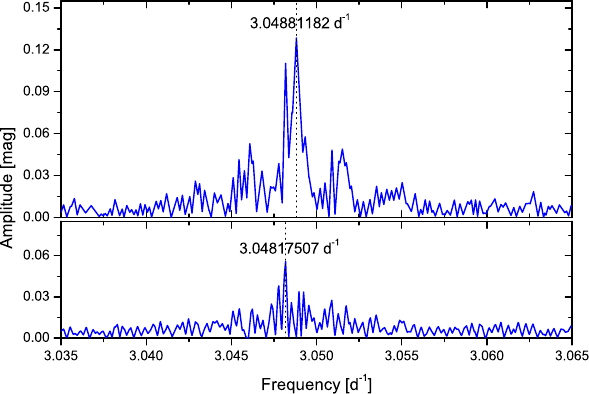}
\caption{Fourier spectrums around the main pulsation frequency based on OGLE I-band data. The sidepeak can be seen clearly.} \label{Fig.9}
\end{figure*}

The period and/or amplitude modulations would cause the sidepeaks around the main frequencies in the Fourier spectra \citep{2011MNRAS.417..974B}. Therefore, the period modulation in BE Dor should also be reflected in the spectrum. However, the data durations of TESS, ASAS-SN, and MACHO projects are shorter than 8 years (2.76, 4.34, and 7.36 years, respectively). The duration of DASCH data is as long as 29 years, but the time resolution is too low. These data are not suitable for finding sidepeaks. Consequently, we use the I-band light curves of OGLE-III and IV for corresponding analysis (see Figure~\ref{Fig.9}). The upper panel in Figure~\ref{Fig.9} shows the Fourier spectrum around the main pulsation frequency (3.04881182 d$^{-1}$), from which the sidepeak phenomenon can be seen clearly. The bottom panel shows the spectrum after the data are pre-whitened with the main frequency and its harmonics, and the peak in the panel is the corresponding single sideband (3.04817507 d$^{-1}$). According to the frequency difference, the corresponding period is about 4.3 years, which is about half of the modulation period. The reason for this result may be related to the abrupt period change occurred between OGLE-III and IV. Based on space and OGLE data, some RRc stars are found to exist additional low-amplitude frequencies $f_{2}$, which have a period ratio of $P_{2}/P_{1}$ = 0.612 - 0.632 \citep{2015MNRAS.447.2348M,2019MNRAS.487.5584N,2021arXiv210907329M}. To explore whether there is the same pulsation component in BE Dor, we use TESS data for spectrum analysis. However, no significant peak is found in the corresponding frequency range. Maybe the accuracy of the data we obtained is not enough to support the discovery of weak periodicities.

\section{Discussion}  \label{Sec:Discussion}

The above analyses show that the pulsation modulation of the pulsating variable BE Dor is mainly manifested in its period modulation, while the variations of the other pulsation parameters and Fourier coefficients, such as the $R_{21}$, $R_{31}$, $\phi_{21}$ and $\phi_{31}$, are relatively weaker. In addition, the $O-C$ diagram (Figure~\ref{Fig.4}) and the $P_{\rm pul}$ - Time diagram (Figure~\ref{Fig.6}) also indicate that the pulsation period shows quasi-periodic variation and abrupt change. Therefore, the light-travel time effect caused by the presence of a companion object cannot explain these variations. According to the definition, BE Dor can be considered as one Blazhko star. However, the variation characteristics (only pulsation period modulation, quasi-periodic variation with long period up to several years) are also different from typical Blazhko RRc stars \citep{2018MNRAS.480.1229N}. The theoretical researches trying to explain Blazhko phenomenon have been widely carried out, and researchers have put forward some different models and viewpoints (see \citealt{2014AJ....148...88C} and references therein). Perhaps one of them can explain the modulation phenomenon of BE Dor.

\subsection{the Stothers' viewpoints}

\citet{2004MNRAS.354..821D} also mentioned another possible mechanism leading to the modulation, i.e., hydromagnetism \citep{1980PASP...92..475S}. \citet{1980PASP...92..475S} pointed out that the hydrogen and helium ionization zones of RRLs are convective unstable, and it can be reasonably assumed that the local turbulent motions can directly interfere with the pulsation of the envelope. However, Stothers expected that the periodic variation of pulsation caused by convection is random and short-term (hours to days), would not accumulate on a long time scale, and the observed pulsation period variations are related to the activity of solar-like magnetic field. Stothers inferred that the sudden change of some pulsation periods may correspond to the sudden generation or destruction of the magnetic field. If so, the quasi-periodic and abrupt changes in BE Dor $O-C$ diagram can be explained by the viewpoint of \citet{1980PASP...92..475S}. However, according to some observational studies \citep{2004A&A...413.1087C,2009A&A...498..543K}, no significant longitudinal magnetic field was found in RRLs.

Subsequently, Stothers further improved the corresponding theoretical viewpoint to explain the Blazhko effect \citep{2006ApJ...652..643S,2010PASP..122..536S,2011PASP..123..127S}. \citet{2006ApJ...652..643S} considered the interaction between turbulent convection and magnetic field activity, and pointed out that the cyclically strengthening and weakening of turbulent convection in the hydrogen and helium ionization zones is the mechanism causing the Blazhko effect. However, through research, \citet{2011MNRAS.414.2950S} and \citet{2012MNRAS.424...31M} found that Stothers model cannot explain Blazhko modulations with short period (less than 100 days). Noting the period modulation characteristics of RRc stars such as BE Dor and those mentioned in Section~\ref{sec:intro}, we suppose that the viewpoints of Stothers may not explain the Blazhko effect, but maybe suitable to explain the period modulation of BE Dor, that is, the strong modulations with a period of several years are caused by the interaction of turbulent convection and magnetic field activities, which affects the pulsation in the coupled state.

\subsection{Four RRc stars in $Kepler$ field}

We note that the energy equipartition and the Equation (1) mentioned in \citet{1980PASP...92..475S}:
\begin{equation}
\frac{H^{2}}{8\pi} \approx \frac{1}{2} \rho \upsilon_{\rm turb}^{2}.
\end{equation}
It can be seen that the magnetic field strength $H$ is positively correlated with the turbulence velocity $\upsilon_{\rm turb}$, and they are interrelated. Therefore, the intensity of $H$ can be known indirectly by detecting the $\upsilon_{\rm turb}$.

The information of $\upsilon_{\rm turb}$ can be obtained from spectral data. However, due to the faint nature of BE Dor, the relevant spectral observations are still few. In Section~\ref{sec:intro} of this paper, we mentioned some RRc stars showing similar period modulations. The variations of BE Dor are not a rare case. Maybe we should look for some more similar samples, get more clues by studying their characteristics.

\begin{figure*}
\centering
\includegraphics[width=.45\textwidth]{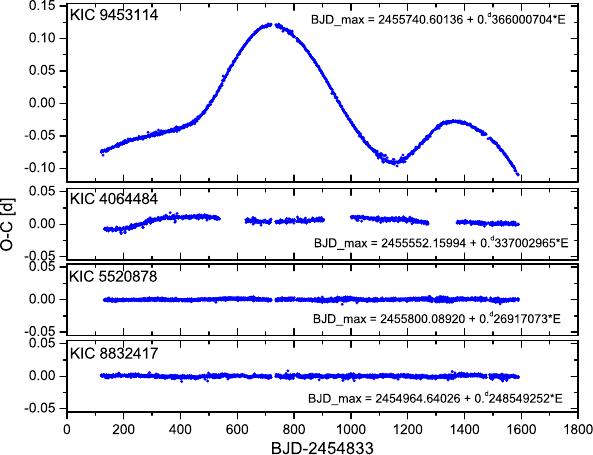}
\caption{$O-C$ diagrams of four RRc stars in $Kepler$ field.} \label{Fig.10}
\end{figure*}

\citet{2013ApJ...773..181N} used the large aperture grounded telescopes to observe the RRLs in the $Kepler$ field, obtained the high-resolution spectra, and determined the corresponding physical characteristics. In their paper, four RRc stars are included. Using the uninterrupted photometric data provided by $Kepler$, we determine their times of light maximum and plot their $O-C$ diagrams (see Figure~\ref{Fig.10}, refer to \citealt{2014MNRAS.444..600L} for the corresponding data processing and analysis methods), in which the $O-C$ diagram for KIC 9453114 shows extremely large variation, similar to the changes of BE Dor, those of KIC 5520878 and 8832417 have little changes and are almost straight lines, and that of KIC 4064484 has weak fluctuation. Table~\ref{Table3} lists their physical parameters provided by \citet{2013ApJ...773..181N} (see Table 9 in their paper), in which the four stars are arranged from left to right according to the degree of $O-C$ changes. It can be found that the parameters of KIC 9453114 showing the largest $O-C$ variation are different from other 3 RRc stars: its effective temperature $T_{\rm eff}$, surface gravity acceleration $\log g$ and metal abundance [M/H] are the smallest, while the macroturbulent velocity $v_{\rm mac}$ and microturbulent velocity $\xi_{t}$ are the largest. Especially, the $v_{\rm mac}$ reaches 29.3 km s$^{-1}$, which is the largest among RRLs in the $Kepler$ field, about 2-3 times that of other stars. The two RRc stars with almost no $O-C$ changes have the highest $T_{\rm eff}$, [M/H] and the smaller $v_{\rm mac}$. Therefore, our direct idea is that there is a positive correlation between the changing severity of $O-C$ diagram and the $v_{\rm mac}$, and the reason for the large $v_{\rm mac}$ may be related to the $T_{\rm eff}$, [M/H] and the implied evolution history.

In Table~\ref{Table3}, it can be found that KIC 9453114 has the longest pulsation period and the lowest $T_{\rm eff}$, $\log g$ and [M/H]. These parameters seem to imply that it is an old and evolved RR Lyrae star, at the late stage of horizontal branch. The variable stars in this evolution stage move to the upper right in the HR diagram, the $T_{\rm eff}$ decrease, and the radii increase (resulting in the decrease of $\log g$). At relatively low temperature, the turbulence develops more fully, which affects the pulsation, resulting in the pulsation period variation on the order of $\delta P/P$ $\sim$ 10$^{-4}$.

Other macroturbulent velocity measurements and research works on RRLs are also being carried out. \citet{2019AJ....157..153P} derived the macroturbulent velocity variations of RRLs, and found that the RRab stars exhibit an upper limit on $\upsilon_{\rm mac}$ of $5\pm1$ km s$^{-1}$, and the RRc minima range from 2 to 12 km s$^{-1}$. Compared with their results, those obtained by \citet{2013ApJ...773..181N} are generally and systematically larger. The reason for this may be related to the use of different analysis roadmaps and methods.

\subsection{More discussions}

Admittedly, due to the lack of spectral data of BE Dor, our view that the pulsation period modulation is related to the macroturbulent velocity is obtained indirectly, which is a preliminary conclusion. More works are needed in the future.

During the review phase of this paper, we noticed that \citet{2021tsc2.confE..53D} have also been paying attention to BE Dor. In their poster, they also used OGLE and TESS data for analysis, confirmed the monoperiodic nature of BE Dor, and also noted that the $O-C$ diagram is no longer periodic. More importantly, they have carried out spectral observation on BE Dor since 2004 and accumulated a large amount of spectral data for determining the radial velocities. These data are very important and valuable. It is worth carrying out in-depth data processing and analysis based on these data to detect more atmospheric physical parameters that may play a key role in the study of period modulations in BE Dor.

On the other hand, find more RRc stars that show similar modulation phenomena, discover their natural features, and provide clues for theoretical work. For example, \citet{2018MNRAS.480.1229N} presented the analyses of Blazhko RRc stars in the Galactic bulge by using OGLE-IV data. Figure 6 in their paper plotted the histogram of Blazhko periods. We note that there is a weak centralized distribution of Blazhko RR stars in the period of 1500 - 3000 days. Perhaps similar period modulations exist in these targets and they can be used as research objects of our next studies.

The purpose of Stothers \citep{1980PASP...92..475S,2006ApJ...652..643S,2010PASP..122..536S,2011PASP..123..127S} was to explain the Blazhko mechanism, but it was also considered that there are some deficiencies in Stothers model \citep{2011MNRAS.414.2950S,2012MNRAS.424...31M}. We suppose that the idea of the interaction between the turbulence and magnetic field activities may cause the phenomena on those RRc stars mentioned in this paper. It should be recognized that it would be extremely challenging to model such interactions, which would need 3D modeling of various processes. The modeling is beyond our ability and scope of this primarily observational paper. Recently, some more complex pulsation models have been proposed to explain the observed phenomena \citep{2011MNRAS.414.1111K}. We hope our findings can provide some information and inspiration for their work.

Our work on BE Dor draws on the observation results of several sky survey projects. Their main scientific objects are not pulsating variables (but the baryonic dark matters, supernovae, and exoplanets). However, they have greatly helped the development of this research field. It is worth mentioning that TESS project, due to its design characteristics (small aperture and low spatial resolution), seems that the faint source in crowded star fields such as BE Dor cannot obtain any pulsation information. Nevertheless, our work shows that as long as targeted data processing is adopted, valuable data and information can be obtained from the original image data.

\section{Summary}  \label{Sec:Summary}

Using the photometric data of several sky surveys, we carried out $O-C$ and Fourier analyses for RRc star BE Dor. Through analyses, we have the following discoveries and viewpoints:

1.	The change in the $O-C$ diagram of BE Dor is no longer the periodic change pointed out by \citet{2004MNRAS.354..821D}, but quasi-periodic change (the period is about 8.1 years) and sudden jump (occurring during the observation period of OGLE project). Therefore, the light-travel time effect caused by the motion of the companion object can no longer explain the change in the $O-C$ diagram.

2.	The light curves provided by MACHO, OGLE, and TESS are Fourier decomposed to obtain the pulsation period, mean magnitude, and Fourier coefficients of BE Dor. The change of the pulsation period with time supports the results of the $O-C$ diagram. Other parameters also fluctuate with time but are weaker than the period modulation. It indicates that BE Dor is a monoperiodic RRc star with significant period modulation.

3.	Such modulations are not rare in RRc stars. Through the comparison of four RRc stars in the $Kepler$ field, we found that one of RRc stars, KIC 9453114, showed strong $O-C$ variations similar to BE Dor, and its physical parameters ($T_{\rm err}$, $\log g$, and $v_{\rm mac}$) are different from other three stars. The most notably is its high $v_{\rm mac}$, up to 29.3 km s$^{-1}$. Moreover, the variation degree of the $O-C$ diagram for the four RRc stars also seems to be related to the $v_{\rm mac}$, i.e., the $O-C$ diagrams do not change for two stars with relative low $v_{\rm mac}$; but changes rapid and dramatic for KIC 9453114.

4.	The period modulations in BE Dor, KIC 9453114 and other RRc stars showing similar phenomena (mentioned in Section~\ref{sec:intro}) can be explained by the Stothers' viewpoint, that is, the interaction between turbulent convection and magnetic field activities in the ionization region leads to the instability of pulsation period \citep{1980PASP...92..475S,2006ApJ...652..643S}. Stothers' viewpoint may not explain the general Blazhko effect, but it may explain the modulation phenomenon of some RRLs.

Our work mainly used photometric data, and the conclusion has some limitations. The spectral observations of pulsating variables that can provide more information are very necessary. Investigators have made many systematic achievements by using spectral data from different sources (\citealt{2021ApJ...909...25D} and references therein). In addition, the future space photometric missions, most prominently ESA's PLATO space mission (PLAnetary Transits and Oscillations of stars mission, \citealt{2016AN....337..961R}), would also play an important role in the follow-up observations of BE Dor and in finding similar systems. It is believed that in the future, with the help of observation and data accumulation, we can more comprehensively understand the properties of RRLs and finally solve the various problems left over from history.

\section{acknowledgments}

This paper utilizes public domain data obtained by the MACHO Project, jointly funded by the US Department of Energy through the University of California, Lawrence Livermore National Laboratory under contract No. W-7405-Eng-48, by the National Science Foundation through the Center for Particle Astrophysics of the University of California under cooperative agreement AST-8809616, and by the Mount Stromlo and Siding Spring Observatory, part of the Australian National University. The DASCH project at Harvard is grateful for partial support from NSF grants AST-0407380, AST-0909073, and AST-1313370. This work is supported by the National Natural Science Foundation of China (No.11803084, No.11933008, and No.11922306).

\section{Date availability statements}

The data used in this article are from five sky survey projects, and the original data or images can be obtained through the following links query:

\url{https://projects.iq.harvard.edu/dasch} (DASCH)

\url{https://mast.stsci.edu/tesscut/}(TESS)

\url{http://www.astronomy.ohio-state.edu/asassn} (ASAS-SN)

\url{https://wwwmacho.anu.edu.au/} (MACHO)

\url{http://ogle.astrouw.edu.pl/} (OGLE)

The data generated by our analyses are available in the article and in its online supplementary material.



\bibliographystyle{mnras}
\bibliography{Ref} 


\begin{table}
\begin{center}
\caption{Times of light maximum and corresponding errors for BE Dor provided by present paper. (the first 10 lines of the whole table)}\label{Table1}
\begin{tabular}{l c c c}
\hline\hline
   Max.	       &	Error	&	Method/Filter	& Ref.	\\
	HJD        &	(days)	&		            &	    \\
\hline
2425921.9326 	&	0.0043 	&	pg	&	DASCH	\\
2426001.9809 	&	0.0043 	&	pg	&	DASCH	\\
2426141.4127 	&	0.0044 	&	pg	&	DASCH	\\
2426256.8961 	&	0.0044 	&	pg	&	DASCH	\\
2426564.3070 	&	0.0044 	&	pg	&	DASCH	\\
2426601.0520 	&	0.0044 	&	pg	&	DASCH	\\
2426820.5241 	&	0.0046 	&	pg	&	DASCH	\\
2426939.9301 	&	0.0046 	&	pg	&	DASCH	\\
2427071.1420 	&	0.0046 	&	pg	&	DASCH	\\
2427358.1483 	&	0.0045 	&	pg	&	DASCH	\\
...             &   ...     &   ... &    ...    \\
\hline\hline
\end{tabular}
\end{center}
{\scriptsize {\bf Notes.} pg - photographic; RI - the wavelength coverage of TESS is 6000 - 10000 angstrom, covering most of the Johnson/Bessell R and I filters. The TESS times are provided in BJD\_TDB. For consistency, we have converted the times in BJD\_TDB to in HJD\_UTC \citep{2010PASP..122..935E}.}
\end{table}

\begin{table*}
\centering
\begin{minipage}{175mm}
\caption{Fourier parameters and coefficients for BE Dor from different data sources. (the first 10 lines of the whole table)}\label{Table2}
{\scriptsize
\def\arraystretch{1.5}
\tabcolsep=1.3pt
\begin{tabular}{l c c c c c c c c}
\hline\hline
Data Set	& Filter&	Mean Time & $A_{0}$   & $P_{\rm pul}$ &	$R_{21}$ & $\phi_{21}$ & $R_{31}$ &	$\phi_{31}$	      \\
	        &		& HJD-2400000 &	 (mag.)	  &	  (days)      &	         &	(rad)	   &          &  	(rad)	      \\
\hline
MACHO\_01	 &   b	&  48906.1891 &	-9.172 &	0.327981(06) &	0.177(13) &	3.149(.077) & 	0.064(12) &	 0.218(.228)  \\
MACHO\_02	 &   b	&  49020.4754 &	-9.166 &	0.327993(10) &	0.164(11) &	3.268(.082) & 	0.094(13) &	 0.078(.120)  \\
MACHO\_03	 &   b	&  49081.0099 &	-9.161 &	0.327992(10) &	0.191(09) &	3.261(.061) & 	0.081(11) &	-0.289(.122)  \\
MACHO\_04	 &   b	&  49144.4675 &	-9.157 &	0.328007(08) &	0.185(13) &	3.157(.069) & 	0.082(13) &	-0.189(.153)  \\
MACHO\_05	 &   b	&  49233.6808 &	-9.166 &	0.327992(05) &	0.198(14) &	3.188(.069) & 	0.091(12) &	-0.206(.161)  \\
MACHO\_06	 &   b	&  49356.5275 &	-9.168 &	0.328006(04) &	0.173(09) &	3.233(.063) & 	0.080(10) &	-0.171(.139)  \\
MACHO\_07	 &   b	&  49460.4636 &	-9.156 &	0.328018(07) &	0.171(12) &	3.116(.075) & 	0.082(13) &	 0.099(.143)  \\
MACHO\_08	 &   b	&  49525.1692 &	-9.156 &	0.328039(11) &	0.149(11) &	3.237(.071) & 	0.087(10) &	-0.088(.123)  \\
MACHO\_09	 &   b	&  49582.6118 &	-9.171 &	0.328017(05) &	0.198(11) &	3.093(.041) & 	0.102(09) &	-0.664(.095)  \\
MACHO\_10	 &   b	&  49641.1920 &	-9.169 &	0.328039(05) &	0.173(10) &	3.251(.058) & 	0.099(09) &	-0.019(.103)  \\
...          &  ... &    ...      &   ...  &       ...       &    ...     &    ...      &     ...     &    ...        \\
\hline\hline
\end{tabular}
}
\end{minipage}
{\scriptsize {\bf Notes.} The mean magnitude $A_{0}$ for TESS data are not real, therefore, are not listed.}
\end{table*}

\begin{table}
\begin{center}
\caption{Physical characteristics of the four RRc stars in $Kepler$ provided by \citet{2013ApJ...773..181N}.}\label{Table3}
\begin{tabular}{l c c c c}
\hline\hline
   KIC Number                 & 9453114 & 4064484 & 5520878 & 8832417 \\
\hline
   $P_{\rm pul}$ (days) 	  &	0.366 	& 0.337	  &	0.269   & 0.248   \\
   $T_{\rm eff}$ (K)          & 6204    & 6473    & 7266    & 6999    \\
   $\log g$ (cm s$^{-2}$)      & 2.04    & 2.39    & 3.66    & 3.30    \\
   $[\rm M/H]$ (dex)          & -2.15   & -1.51   & -0.18   & -0.25   \\
   $v_{\rm mac}$ (km s$^{-1}$)& 29.3    & 18.4    & 15.2    & 15.3    \\
   $\xi_{t}$ (km s$^{-1}$)& 6.4     & 4.0     & 3.0     & 3.5     \\
   $v \sin i$ (km s$^{-1}$)   & 4.3     & 4.9     & 5.1     & 4.8     \\
\hline\hline
\end{tabular}
\end{center}
\end{table}

\bsp	
\label{lastpage}
\end{document}